\documentclass[a4paper,12pt]{article}
\usepackage{amsmath}
\usepackage{amssymb}
\usepackage{amsmath,amsthm,amssymb,amscd}
\usepackage{latexsym}
\usepackage{indentfirst}
\usepackage{subfigure}
\usepackage{graphicx}
\usepackage[top=1in,bottom=1in,left=1.25in,right=1.25in]{geometry}

\makeatother
\date{}
\theoremstyle{plain}

\begin{document}

\bibliographystyle{unsrt}
\bibliographystyle{plain}

\title{
  \bf The efficiency of molecular motors
}

\author{{Yunxin Zhang}\thanks{
School of Mathematical Sciences, Fudan University,  Shanghai 200433,
China}
\thanks{Shanghai Key Laboratory for Contemporary Applied Mathematics, Fudan
University}
\thanks{Centre for Computational Systems Biology, Fudan University   (E-Mail: xyz@fudan.edu.cn)}
}

\maketitle \baselineskip=6mm

\baselineskip=22pt

\begin{abstract}
Molecular motors convert chemical energy into mechanical work while
operating in an environment dominated by Brownian motion. The aim of
this paper is to explore the flow of energy between the molecular
motors and its surroundings, in particular, its efficiency. Based on
the Fokker-Planck equation with either $N$ or infinite chemical
states, we find that the energy efficiency of molecular motors,
whether the Stokes efficiency or the usual thermodynamic efficiency,
is strictly less than one, because of the dissipation of the energy
in both the overdamped surroundings and in the process of the
chemical reaction.

\vspace{2em} \noindent \textit{PACS}: 87.16.Nn, 87.16.A-, 82.39.-k,
05.40.Jc

\vspace{2em} \noindent \textit{Keywords}: molecular motors;
Fokker-Planck equation; energy efficiency
\end{abstract}

\section{Introduction}
Molecular motors are biogenic force generators acting in the
nanometer range (\cite{Howard2001}). Linear motors produce sliding
movements along filamentous structures called protein tracks; for
example, myosin slides along an actin filament(\cite{Christof2006,
Anneka2007, Katsuyuki2007}), kinesin ( \cite{Block1990, Carter2005,
David2007, Yuichi2005, Vale1996}) and dynein (\cite{Sakakibara1999,
Vale2003} ) along a microtubule. Many of molecular motors have the
ability to rotate; for example, the bacterial flagellar motor
(\cite{Kudo1990}) and $F_0F_1$-ATP synthase (\cite{Noji1997}). This
movements and forces generated by the motors play essential roles in
cellular functions. The linear motor kinesins are widely distributed
in almost all eukaryotic cells. These motors are adenosine
triphosphate (ATP)-driven walking machines that move in 8-nm steps
toward the plus ends of the microtubule, turn over one ATP molecular
per step under a range of loads (\cite{Yildiz2004, Asbury2003,
Howard2001, Masayoshi2002} ). Kinesin has two motor domains called
heads, each of which includes one ATP- and microtubule binding site
( \cite{Maria2007, Kull1996}). This motor steps mainly in the
forward direction (to the plus end of the microtubule), but
occasionally in the backward direction (to its minus end)
(\cite{voboda1994}).

The study of the energetics of molecular motors is relevant for
several reasons. Firstly, the efficiency of molecule motors should
be high enough to decrease the heat dissipation. Otherwise, the
temperature of the surrounding would become higher and higher since
the heat is difficult to be dissipated. Secondly, molecular motors
are related to fundamental problems of thermodynamics and
statistical mechanics. For instance, one of the first well-known
Brownian motors was the ratchet-and-pawl setup studied by Feynman in
his Lectures \cite{Feynman1963}. He calculated the efficiency of
such an engine since his original purpose was to prove that an
automatic demon cannot beat the second law of thermodynamics.
Thirdly, many models of molecular motors proposed in the literature
are based on nonequilibrium fluctuations without specifying their
source. However, the study of the energetics of such models requires
a more precise formulation, since one has to determine the physical
nature of the external agent and verify that the motor is consistent
with the second law of thermodynamics (\cite{parrondo, Qian2004}).

Recently, Hongyun Wang and G. Oster \cite{Wang2002} have given the
definition of the Stokes efficiency $\eta_S$ (see Eq. (\ref{16})),
and shown $\eta_S\le 1$ through mathematical analysis of the
Fokker-Planck equation with $N$ chemical states. In \cite{Qian2004},
Hong Qian have introduced a nonequilibrium potential function for a
motor protein modelled by a rectified Brownian motion. Since the
conservation of energy can be mathematically established, Qian's
model provides a comprehensive theory for motor efficiency. In
\cite{parrondo}, the energetics of forced ratchets, chemical motors,
and thermal motors has been discussed. The reasonable definitions of
the energy efficiency of motors, which are compatible with the law
of the thermodynamics, have also been reviewed.

In this research, we will discuss some properties of the energy
efficiency of molecular motors in more details. Using the
description of the Fokker-Planck equation with $N$ or infinite
chemical states, we obtain $\eta_S< 1, \eta_T< 1$. It is to say that
the energy efficiency of molecular motors, either the Stokes
efficiency (\cite{Wang2002}) or the usual thermodynamic efficiency
(\cite{parrondo, Qian2004}), is strictly less than one, and any
attempt to design a molecular motor that can convert the chemical
energy into mechanical energy with efficiency one would fail. The
movement of molecular motors satisfies the second law of the
thermodynamics. It is why the Feynman ratchet doesn't work
\cite{Reimann2002, Feynman1963}. But how is the chemical energy
consumed in the movement of the molecular motors? In this paper,
we'll find that the total energy released by the fuel molecules is
consumed in three ways: to do mechanical work, to be dissipated into
the surroundings, to be dissipated in the chemical reaction (see Eq.
(\ref{40})). In fact, for motor proteins, such as kinesin, myosin
and dynein, the heat dissipation in the surroundings is very
important to keep the body temperature and a well-balanced body
temperature is essential to the functions of the cell. So, not only
mathematical analysis but also biophysical demand implies that the
energy efficiency of the molecular motor should be strictly less
than one. For motor proteins, the chemical energy is not only used
to do mechanical work but also used to keep the body temperature. Of
course, high efficient molecular motors are also very desirable in
mechanical equipment.

\section{Fokker-Planck equation with $N$ chemical states}
In this section, we'll discuss the energy efficiency $\eta_S$ and
$\eta_T$ using the Fokker-Planck equation with $N$ chemical states.

Let $\rho_i(x, t)$ be the probability density for finding the motor
at position $x$ in the $i-$th chemical state at time $t$. The
evolution of the probability density is governed by a set of coupled
Fokker-Planck equations that ensure the conservation of probability:
\begin{equation}\label{1}
\frac{\partial \rho_i(x,t)}{\partial t}+\frac{\partial
J_i(x,t)}{\partial x}=I_{i-\frac12}(x,t)-I_{i+\frac12}(x,t)\qquad
1\le i\le N
\end{equation}
Here $J_i$ is the probability flux in the spatial direction for the
$i-$th chemical state, associated with convective transport and
diffusion. $I_{i+1/2}$ is the probability flux density from state
$i$ to state $i + 1$, $N$ is the number of chemical states in one
reaction cycle. For example, $N=4$ is usually used in the motility
cycle of ATP-driven motor proteins (\cite{Vale2000, Wang2002}).

In the $i-$th state, the motor is driven by a periodic potential
$\phi_i(x)$, and
\begin{equation}\label{2}
\begin{aligned}
J_i(x,t)&=\rho_i(x,t)u_i(x,t)=-\frac{\rho_i(x,t)}{\xi}\frac{\partial
\Phi_i(x,t)}{\partial x}\cr &=-\frac{\rho_i(x,t)}{\xi}\frac{\partial
\bar{\phi}_i(x)}{\partial x}-D\frac{\partial \rho_i(x,t)}{\partial
x}\cr &=\underbrace{-\frac{\rho_i(x,t)}{\xi}\frac{\partial
{\phi}_i(x)}{\partial
x}+\frac{\rho_i(x,t)f_{ext}}{\xi}}\limits_{spatial\ drift\
flux}\underbrace{-D\frac{\partial \rho_i(x,t)}{\partial
x}}\limits_{spatial\ diffusion\ flux}
\end{aligned}
\end{equation}
\begin{equation}\label{3}
I_{i+\frac12}(x,t)=\rho_i(x,t)k^+_i-\rho_{i+1}(x,t)k^-_{i+1}\
(\textrm{flux along reaction coordinates})
\end{equation}
where $\Phi_i=\bar{\phi}_i+k_BT\ln \rho_i=\phi_i-f_{ext}x+k_BT\ln
\rho_i$ is the enthalpy, $\xi$ is the coefficient of viscous drag,
$D$ is the free diffusion coefficient, $k_B$ is the Boltzmann
constant, $T$ is absolute temperature ($D, \xi, k_B, T$ satisfy the
Einstein relation $k_BT=\xi D$), $f_{ext}$ is the external load
($f_{ext}\le 0$ ), and $k^+_i$ is the transition rate from state $i$
to state $i+1$, $k^-_{i+1}$ is the transition rate from state $i+1$
to state $i$. $k^+_i$ and $k^-_{i+1}$ satisfy
\begin{equation}\label{4}
\frac{k^+_i}{k^-_{i+1}}=\exp\left(\frac{\phi_i(x)-\phi_{i+1}(x)}{k_BT}\right)
\end{equation}
which ensures detailed balance at equilibrium.

Since the motor operates in a chemical and mechanical cycle, the
boundary conditions for Eq. (\ref{1}) in both the spatial direction
and the reaction coordinate are periodical: $\rho_i(x+L,
t)=\rho_i(x, t)$, $\rho_{i+N}(x, t)=\rho_i(x, t)$,
$\phi_{i+N}=\phi_i-\triangle\mu$, where $L$ is the step size of the
motor ($L=8.2$ nm for motor protein kinesin), $\triangle\mu>0$ is
the chemical free energy consumed in one reaction cycle. In
physiological conditions, the hydrolysis free energy of ATP is
$\triangle\mu\thickapprox 25k_BT$ (\cite{Howard2001, Zhang2008,
Wang2007}).

At steady state, the dissipation rate $\Pi_i$ corresponding to slide
within the potential profile $\Phi_i$ (i.e. the dissipation rate in
the $i-$th chemical state) is (\cite{Groot, Parmeggiani})
\begin{equation}\label{5}
\Pi_i=-\int_0^LJ_i\frac{\partial \Phi_i}{\partial x}dx\qquad 1\le
i\le N
\end{equation}
It can be readily verified that
$\Pi_i=\int_0^L\frac{\rho(x)}{\xi}\left(\frac{\partial
\Phi_i}{\partial x}\right)^2dx\ge 0$. Summing (\ref{5}) over $i$ and
integral by part, we obtain the total spatial dissipation rate
\begin{equation}\label{6}
\begin{aligned}
\Pi=&\sum_{i=1}^N\Pi_i=-\int_0^L\sum_{i=1}^NJ_i\frac{\partial
\Phi_i}{\partial x}dx\cr
=&f_{ext}L\sum_{i=1}^NJ_i(0)+\int_0^L\sum_{i=1}^N\frac{J_i(x)}{\partial
x}\Phi_i(x)dx
\end{aligned}
\end{equation}
In view of Eq. (\ref{1}), at steady state,
\begin{equation}\label{61}
\frac{\partial J_i(x)}{\partial x}=I_{i-\frac12}(x)-I_{i+\frac12}(x)
\end{equation}
Eqs. (\ref{6}) (\ref{61}) imply
\begin{equation}\label{7}
\begin{aligned}
\Pi=&f_{ext}L\sum_{i=1}^NJ_i(0)+\int_0^L\sum_{i=1}^N(I_{i-\frac12}(x)-I_{i+\frac12}(x))\Phi_i(x)dx\cr
=&f_{ext}L\sum_{i=1}^NJ_i(0)+\int_0^L\sum_{i=1}^{N}(\Phi_{i+1}(x)-\Phi_i(x))I_{i+\frac12}(x)dx+\int_0^LI_{\frac12}(x)(\Phi_1(x)-\Phi_{N+1}(x))dx\cr
=&f_{ext}L\sum_{i=1}^NJ_i(0)+\triangle\mu\int_0^LI_{\frac12}(x)dx+\int_0^L\sum_{i=1}^{N}(\Phi_{i+1}(x)-\Phi_i(x))I_{i+\frac12}(x)dx
\end{aligned}
\end{equation}
By Eqs. (\ref{6}) and (\ref{7}), we obtain
\begin{equation}\label{8}
f_{ext}L\sum_{i=1}^NJ_i(0)+\triangle\mu\int_0^LI_{\frac12}(x)dx=\sum_{i=1}^N\int_0^L\left[\frac{\rho(x)}{\xi}\left(\frac{\partial
\Phi_i}{\partial
x}\right)^2dx+(\Phi_{i}(x)-\Phi_{i+1}(x))I_{i+\frac12}(x)\right]dx
\end{equation}

Thanks to Eq. (\ref{61}),
\begin{equation}\label{9}
\frac{\partial}{\partial x}\left(\sum_{i=1}^NJ_i(x)\right)\equiv
0\quad
\int_0^LI_{\frac12}(x)dx=\int_0^LI_{\frac32}(x)dx=\cdots=\int_0^LI_{N+\frac12}(x)dx
\end{equation}
it is to say
\begin{equation}\label{10}
L\sum_{i=1}^NJ_i(0)=\sum_{i=1}^N\int_0^LJ_i(x)dx=\int_0^L\sum_{i=1}^N\rho_i(x)u_i(x)dx=:V
\end{equation}
\begin{equation}\label{11}
\int_0^LI_{\frac12}(x)dx=\frac1N\sum_{i=1}^N\int_0^LI_{i+\frac12}(x)dx=:\nu
\end{equation}
where $V$ is the mean velocity of molecular motors, $\nu$ is the
mean rate of the chemical reaction.

Due to (\ref{3}) (\ref{4})
\begin{equation}\label{12}
\begin{aligned}
I_{i+\frac12}(x,t)=&\rho_i(x,t)k^+_i\left[1-\frac{\rho_{i+1}(x,t)k^-_{i+1}}{\rho_i(x,t)k^+_i}\right]\cr
=&\rho_i(x,t)k^+_i\left[1-\frac{\rho_{i+1}}{\rho_i(x,t)}\exp\left(\frac{\phi_{i+1}(x)-\phi_i(x)}{k_BT}\right)\right]\cr
=&\rho_i(x,t)k^+_i\left[1-\exp\left(\frac{\bar{\phi}_{i+1}(x)-\bar{\phi}_i(x)}{k_BT}\right)\right]\cr
=&\rho_i(x,t)k^+_i\left[1-\exp\left(\frac{{\Phi}_{i+1}(x)-{\Phi}_i(x)}{k_BT}\right)\right]
\end{aligned}
\end{equation}
so
\begin{equation}\label{13}
\begin{aligned}
&{\Phi}_{i+1}(x)-{\Phi}_i(x)\ge 0 (\le 0)\Longrightarrow
\exp\left(\frac{{\Phi}_{i+1}(x)-{\Phi}_i(x)}{k_BT}\right)\ge 1 (\le
1)\cr \Longrightarrow &I_{i+\frac12}(x,t)\le 0 (\ge
0)\Longrightarrow (\Phi_{i}(x)-\Phi_{i+1}(x))I_{i+\frac12}\ge0
\end{aligned}
\end{equation}

Finally, Eqs. (\ref{8}) (\ref{10}) (\ref{11}) (\ref{13}) mean
\begin{equation}\label{14}
\begin{aligned}
f_{ext}V+\nu\triangle\mu=&\sum_{i=1}^N\int_0^L\left[\frac{\rho(x)}{\xi}\left(\frac{\partial
\Phi_i}{\partial
x}\right)^2+(\Phi_{i}(x)-\Phi_{i+1}(x))I_{i+\frac12}(x)\right]dx\cr
=&\sum_{i=1}^N\int_0^L\left[\underbrace{{\xi}{\rho(x)}u^2_i(x)}\limits_{I}+\underbrace{(\Phi_{i}(x)-\Phi_{i+1}(x))I_{i+\frac12}(x)}\limits_{II}\right]dx\ge0
\end{aligned}
\end{equation}
where $I$ is the dissipation corresponding to slide within the
potential profile $\Phi_i$, $II$ is the dissipation corresponding to
transitions between two chemical states. (\ref{14}) is the second
law of the thermodynamics.

By Eq. (\ref{14}), the thermodynamic efficiency
\begin{equation}\label{15}
\eta_T:=-\frac{f_{ext}V}{\nu\triangle\mu}\le 1
\end{equation}
the Stokes efficiency
\begin{equation}\label{16}
\begin{aligned}
\eta_S:=&\frac{\xi V^2}{f_{ext}V+\nu\triangle\mu}=\frac{\xi
\left(\int_0^L\sum_{i=1}^N\rho_i(x)u_i(x)dx\right)^2}{f_{ext}V+\nu\triangle\mu}\cr
\le &\frac{\xi
\int_0^L\sum_{i=1}^N\rho_i(x)u_i^2(x)dx}{f_{ext}V+\nu\triangle\mu}=\frac{\sum_{i=1}^N\int_0^L\frac{\rho(x)}{\xi}\left(\frac{\partial
\Phi_i}{\partial x}\right)^2dx}{f_{ext}V+\nu\triangle\mu}\cr\le &1
\end{aligned}
\end{equation}
In view of (\ref{13})
\begin{equation}\label{17}
\begin{aligned}
&\sum_{i=1}^N\int_0^L(\Phi_{i}(x)-\Phi_{i+1}(x))I_{i+\frac12}(x)dx=0&\cr\Longleftrightarrow&(\Phi_{i}(x)-\Phi_{i+1}(x))I_{i+\frac12}(x)\equiv0\quad
&\forall\ 1\le i\le N\quad 0\le x\le L \cr
\Longleftrightarrow&\Phi_{i}(x)\equiv\Phi_{i+1}(x)\quad &\forall\
1\le i\le N \quad 0\le x\le L
\cr\Longleftrightarrow&\Phi_{i}(x)\equiv\Phi_{j}(x)\quad &\forall\
1\le i, \ j\le N\quad 0\le x\le L \cr \Longrightarrow&\triangle
\mu=0
\end{aligned}
\end{equation}
Therefore, if $\triangle \mu\ne 0$, then
$\sum_{i=1}^N\int_0^L(\Phi_{i}(x)-\Phi_{i+1}(x))I_{i+\frac12}(x)dx\ne
0$, which implies (see the inequality (\ref{14}) and the definitions
(\ref{15}) (\ref{16}))
\begin{equation}\label{18}
\eta_T< 1\qquad \eta_S< 1
\end{equation}

If $\triangle \mu=0$, then the mean velocity $V\le 0$ (see Eq.
(\ref{14})). It is to say that no useful mechanical work can be done
by molecular motors if there is no input of the energy. If the mean
velocity $V> 0$, i.e., molecular motors do useful work against the
external load $f_{ext}<0$, then $\triangle \mu\ne 0$, which implies
$\sum_{i=1}^N\int_0^L(\Phi_{i}(x)-\Phi_{i+1}(x))I_{i+\frac12}(x)dx\ne
0$, consequently $\eta_T< 1, \eta_S< 1$ (see Eqs.
(\ref{14}-\ref{17})). Therefore, there always exists energy
dissipation as long as the input energy $\triangle \mu\ne 0$.

\section{Fokker-Planck equation with continuous chemical coordinates}
In this section, we'll discuss the energy efficiency $\eta_S$ and
$\eta_T$ using the Fokker-Planck equation with infinite chemical
states.

Following Refs. \cite{Keller2000, Qian2000}, we denote $y$ as the
coordinate of the internal conformational space of the motor
protein, and $x$ as its center of mass with respect to the linear
track. Let $\rho(x, y, t)$ be the probability density for finding
the motor at position $x$ and chemical coordinate $y$ at time $t$,
$\Phi(x,y)$ be the potential. Due to (\ref{12}), in the large $N$
limit,
\begin{equation}\label{19}
I_{i-\frac12}(x)-I_{i+\frac12}(x)=-\frac{\partial}{\partial
y}\left[\frac{1}{\zeta}\rho(x,y,t)\frac{\partial \Phi(x,y)}{\partial
y}\right]=-\frac{\partial}{\partial
y}\left[\frac{D_y}{k_BT}\rho(x,y,t)\frac{\partial
\Phi(x,y)}{\partial y}\right]
\end{equation}
where $D_y=\lim\limits_{N\to \infty}\frac{k^+(y\to
y+\frac1N)}{N^2}$ and $\zeta=k_BT/D_y$. Let
$u(x,y)=-\frac{1}{\xi}\frac{\partial \Phi(x,y)}{\partial x}$,
$v(x,y)=-\frac{1}{\zeta}\frac{\partial \Phi(x,y)}{\partial y}$, in
the large $N$ limit, Eq. (\ref{1}) arrives
\begin{equation}\label{20}
\frac{\partial \rho(x,y,t)}{\partial t}+\frac{\partial
J(x,y,t)}{\partial x}+\frac{\partial I(x,y,t)}{\partial y}=0\qquad
0\le x\le L,\quad 0\le y\le 1
\end{equation}
where $J=\rho u$, $I=\rho v$.

At the steady state, the mean velocity of the movement is
\begin{equation}\label{21}
V=<u>=\int_0^1\int_0^L\rho(x,y)u(x,y)dxdy=\int_0^1\int_0^LJ(x,y)dxdy
\end{equation}
the mean rate of the chemical reaction is
\begin{equation}\label{22}
\nu=<v>=\int_0^1\int_0^L\rho(x,y)v(x,y)dxdy=\int_0^1\int_0^LI(x,y)dxdy
\end{equation}
Define
\begin{equation}\label{23}
\Pi=\int_{0}^L\int_0^1\left[\underbrace{-J(x,y)\frac{\partial
\Phi(x,y)}{\partial x}}\limits_{I'}\underbrace{-I(x,y)\frac{\partial
\Phi(x,y)}{\partial y}}\limits_{II'}\right]dydx
\end{equation}
where $I'$ is the dissipation rate along the spatial coordinate $x$,
$II'$ is the dissipation rate along the reaction coordination $y$.
So $\Pi$ is the total dissipation rate of molecular motors.

It is easy to find
\begin{equation}\label{24}
\Pi=\int_0^1\int_0^L\rho\left(
\frac{\Phi_x^2}{\xi}+\frac{\Phi^2_y}{\zeta}\right)dxdy=\int_0^1\int_0^L\rho\left(
\xi u^2+\zeta v^2\right)dxdy\ge 0
\end{equation}
at the same time
\begin{equation}\label{25}
\begin{array}{rl}
&-\int_0^1\int_0^L J(x,y)\Phi_x(x,y)dxdy\cr
=&-\int_0^1\left[J(x,y)\Phi(x,y)\left.\right|_{x=0}^{x=L}-\int_0^LJ_x(x,y)\Phi(x,y)dx\right]dy\cr
=&f_{ext}L\int_0^1J(L,y)dy+\int_0^1\int_0^LJ_x(x,y)\Phi(x,y)dxdy\cr
=&f_{ext}L\int_0^1J(L,y)dy-\int_0^1\int_0^LI_y(x,y)\Phi(x,y)dxdy\cr
=&f_{ext}L\int_0^1J(L,y)dy-\int_0^L\left[I(x,y)\Phi(x,y)\left.\right|_{y=0}^{y=1}-\int_0^1I(x,y)\Phi_y(x,y)dy\right]dx\cr
=&f_{ext}L\int_0^1J(L,y)dy+\triangle\mu\int_0^LI(x,1)dx+\int_0^1\int_0^LI(x,y)\Phi_y(x,y)dxdy
\end{array}
\end{equation}
so
\begin{equation}\label{26}
\Pi=f_{ext}L\int_0^1J(L,y)dy+\triangle\mu\int_0^LI(x,1)dx
\end{equation}
Since
\begin{equation}\label{27}
\frac{\partial}{\partial
x}\left[\int_0^1J(x,y)dy\right]=\int_0^1\frac{\partial}{\partial
x}J(x,y)dy=-\int_0^1\frac{\partial}{\partial y}I(x,y)dy=0
\end{equation}
\begin{equation}\label{28}
\frac{\partial}{\partial
y}\left[\int_0^LI(x,y)dx\right]=\int_0^L\frac{\partial}{\partial
y}I(x,y)dx=-\int_0^L\frac{\partial}{\partial x}J(x,y)dx=0
\end{equation}
one knows
\begin{equation}\label{29}
\int_0^1J(x,y)dy=\int_0^1J(L,y)dy\qquad\int_0^LI(x,y)dx=\int_0^LI(x,1)dx
\end{equation}
therefore
\begin{equation}\label{30}
L\int_0^1J(L,y)dy=\int_0^L\int_0^1J(x,y)dydx=V
\end{equation}
\begin{equation}\label{31}
\int_0^LI(x,1)dx=\int_0^1\int_0^LI(x,y)dxdy=\nu
\end{equation}

Finally, (\ref{24}) (\ref{26}) (\ref{30}) (\ref{31}) imply
\begin{equation}\label{32}
\Pi=f_{ext}V+\nu\triangle\mu=\int_0^1\int_0^L\rho\left(\xi u^2+\zeta
v^2\right)dxdy\ge 0
\end{equation}
this is the second law of the thermodynamics. By (\ref{32}), the
thermodynamic efficiency
\begin{equation}\label{33}
\eta_T=-\frac{f_{ext}V}{\nu\triangle\mu}=1-\frac{\Pi}{\nu\triangle\mu}\le
1
\end{equation}
the Stokes efficiency
\begin{equation}\label{34}
\begin{aligned}
\eta_S=&\frac{\xi V^2}{f_{ext}V+\nu\triangle\mu}=\frac{\xi
\left(\int_0^L\int_0^1\rho(x,y)u(x,y)dydx\right)^2}{f_{ext}V+\nu\triangle\mu}\cr
\le &\frac{\xi \int_0^L\int_0^1
\rho(x,y)u^2(x,y)dydx}{f_{ext}V+\nu\triangle\mu}\cr =&1-\frac{\zeta
\int_0^L\int_0^1 \rho(x,y)v^2(x,y)dydx}{f_{ext}V+\nu\triangle\mu}\le
1
\end{aligned}
\end{equation}
Similar as (\ref{17})
\begin{equation}\label{35}
\begin{aligned}
&\zeta \int_0^L\int_0^1
\rho(x,y)v^2(x,y)dydx=0&\cr\Longleftrightarrow &v(x,y)=0\qquad &0\le
x\le L,\quad 0\le y\le 1\ \ \ \cr
\Longleftrightarrow&\Phi(x,y)=\Phi(x,z)&0\le x\le L,\quad 0\le y,
z\le 1\cr\Longrightarrow&\triangle\mu=\Phi(x,0)-\Phi(x,1)=0
\end{aligned}
\end{equation}
So, if $\triangle\mu\ne 0$, then $\zeta \int_0^L\int_0^1
\rho(x,y)v^2(x,y)dydx\ne 0$, which means the inequalities (\ref{18})
hold true.

Moreover, applying the Cauchy-Schwarz inequality, and using the
constraint $\int_0^L\int_0^1 \rho(x,y)dydx=1$, one obtains
\begin{equation}\label{36}
\begin{aligned}
&\int_0^L\int_0^1 \rho(x,y)u^2(x,y)dydx\ge
\left(\int_0^L\int_0^1\rho(x,y)u(x,y)dydx\right)^2=V^2\cr
&\int_0^L\int_0^1 \rho(x,y)v^2(x,y)dydx\ge
\left(\int_0^L\int_0^1\rho(x,y)v(x,y)dydx\right)^2=\nu^2
\end{aligned}
\end{equation}
so
\begin{equation}\label{37}
\begin{aligned}
\eta_T=&-\frac{f_{ext}V}{\nu\triangle\mu}\le 1-\frac{\xi
V^2+\zeta\nu^2}{\nu\triangle \mu}=1-\frac{(\xi
L^2+\zeta)V}{\triangle \mu L}
\end{aligned}
\end{equation}
\begin{equation}\label{38}
\begin{aligned}
\eta_S=\frac{\xi V^2}{f_{ext}V+\nu\triangle\mu}\le\frac{\xi V^2}{\xi
V^2+\zeta \nu^2}=\frac{\xi L^2}{\xi L^2+\zeta}=1-\frac{\zeta}{\xi
L^2+\zeta}
\end{aligned}
\end{equation}
here the equalities hold true if and only if $u(x,y)\equiv V$,
$v(x,y)\equiv\nu$. Obviously $\eta_T<1,\ \eta_S<1$.

Finally, by Eq. (\ref{32}) and the definitions of the energy
efficient $\eta_S, \eta_T$, we can get the following interesting
relationship
\begin{equation}\label{39}
\begin{aligned}
\xi V^2=\eta_S(1-\eta_T)\nu\triangle\mu
\end{aligned}
\end{equation}
\begin{equation}\label{40}
\begin{array}{clcccccl}
\triangle\mu&=&-f_{ext}L&+&\xi
VL&+&\left[\frac{L}{V}\int_0^1\int_0^L\rho\left( \xi u^2+\zeta
v^2\right)dxdy-\xi VL\right]&\cr
&=&\eta_T\triangle\mu&+&\eta_S(1-\eta_T)\triangle\mu&+&(1-\eta_S)(1-\eta_T)\triangle\mu&
\end{array}
\end{equation}
From Eq. (\ref{40}), we can know how the energy is consumed in the
movement of molecular motors: where $\eta_T\triangle\mu=f_{ext}L$ is
the energy used to do useful work against the external force
$f_{ext}$, $(1-\eta_T)\triangle\mu$ is the total energy dissipation,
$\eta_S(1-\eta_T)\triangle\mu$ is the energy dissipation due to the
overdamping surroundings, $(1-\eta_S)(1-\eta_T)\triangle\mu$ is the
energy dissipation during the chemical reaction.

Though the 2D Fokker-Planck equation (\ref{20}) is the large $N$
limit of the equations (\ref{1}), most of the results in section 3
can't be obtained directly by the limitation of the corresponding
ones in section 2 (for $\eta=\lim_{n\to +\infty}\eta_n$ and
$\eta_n<1$, it is not always correct that $\eta<1$). Moreover,
through the detailed discussion in section 3, the relationship
between the parameters in equations (\ref{1}) and the parameters in
(\ref{20}) is clarified, and how the energy is consumed in the
movement of the molecular motor is also clarified easily. The
equations (\ref{1}) might be more physical, but it might be more
convenient to get detailed results using equation (\ref{20}).

\section{Conclusions}
In this paper, we have given a detailed discussion of the energy
efficiency of molecular motors. We find that the energy efficiency
of molecular motors, whether the usual thermodynamic efficiency or
Stokes efficiency, is strictly less than one. This result is sharper
than in previous work (\cite{Qian2004, Wang2002}), where $\eta_S\le
1$ and $\eta_T\le 1$ were proved. In each step of a molecular motor,
the total energy released by the fuel molecules is consumed in three
ways: to do mechanical work, to be dissipated in the surroundings,
or to be dissipated in the chemical reaction. According to the
second law of the thermodynamics (\ref{32}), the faster the
molecular motors move, the more the energy is dissipated.
Theoretically, the highest energy efficiency only can be achieved
when the movement of the molecular is a quasi-static process, which
is more like the Carnot heat engine.

For the multiple pathways case (\cite{Astumian2005, Liepelt2007,
Linden2007} ), the results in this paper are also correct. The only
difference is that the energy efficiency (or the energy used to do
mechanical work, the velocity) in different pathways might be
different. Regardless of the pathways, the efficiency of molecular
motors is always strictly less than one.

Finally, similar theoretical analysis can be used to describe the
backward stepping case of motor protein kinesin (\cite{Carter2005}).
If fuel molecules are resynthesized in the backward stepping
process, then $f_{ext}<0, V<0, \triangle\mu<0$ and
$f_{ext}V+\nu\triangle\mu>0$. So the energy efficiency of fuel
molecules resynthesizing $\eta=-\frac{\nu\triangle\mu}{f_{ext}V}$ is
strictly less than one. Generally, in each backward step of a
molecular motor, the total external energy ($f_{ext}V$) is consumed
in three ways: to resynthesize fuel molecules ($-\nu\triangle\mu$),
to be dissipated in the surroundings ($\int_0^1\int_0^L\rho\xi
u^2dxdy$), or to be dissipated in the chemical reaction
($\int_0^1\int_0^L\rho\zeta v^2dxdy$):
$f_{ext}V=-\nu\triangle\mu+\int_0^1\int_0^L\rho\left(\xi u^2+\zeta
v^2\right)dxdy$. Where $\triangle\mu<0$ means fuel molecules are
resynthesized, and $\triangle\mu>0$ means fuel molecules are
consumed.

\vskip 0.5cm

\noindent{\bf Acknowledgments}

The author thanks the reviewers for their help to improve this
paper. This work was funded by National Natural Science Foundation
of China (Grant No. 10701029).


\begin{thebibliography}{10}

\bibitem{Maria2007}
Maria~C. Alonso, Douglas~R. Drummond, Susan Kain, Julia Hoeng, Linda
Amos, and
  Robert~A. Cross.
\newblock An ATP gate controls tibulin binding by the tethered head of
  kinesin-1.
\newblock {\em Science}, 316:120--123, 2007.


\bibitem{Astumian2005}
R~Dean Astumian.
\newblock Biasing the random walk of a molecular motor.
\newblock {\em J. Phys.: Condens. Matter}, 17:S3753--S3766, 2005.

\bibitem{Yildiz2004}
A.Yildiz, M.Tomishige, R.D.Vale, and P.R.Selvin.
\newblock Kinesin walks hand-over-hand.
\newblock {\em Science}, 303:676, 2004.

\bibitem{Block1990}
Block, S.M., Goldstein, L.S.B., and B.J. Schnapp.
\newblock Bead movement by single kinesin molecules studied with optical
  tweezers.
\newblock {\em Nature}, 348:348--352, 1990.

\bibitem{BUSTAMANTE}
Carlos Bustamante, David Keller, and George Oster.
\newblock The physics of molecular motors.
\newblock {\em Acc. Chem. Res.}, 34:412--420, 2001.

\bibitem{Carter2005}
N.~J. Carter and R.~A. Cross.
\newblock Mechanics of the kinesin step.
\newblock {\em Nature}, 435:308--312, 2005.

\bibitem{Christof2006}
J.~Christof, M.~Gebhardt, Anabel E.-M. Clemen, Johann Jaud, and
Matthias Rief.
\newblock Myosin-v is a mechanical ratchet.
\newblock {\em PNAS}, 103:8680--8685, 2006.

\bibitem{Asbury2003}
C.L.Asbury, A.N.Fehr, and S.M.Block.
\newblock Kinesin moves by an asymmetric hand-over-hand mechanism.
\newblock {\em Science}, 302:2130--2134, 2003.

\bibitem{Groot}
S.~R.~De Groot and P.~Mazur.
\newblock {\em Non-Equilibrium Thermodynamics}.
\newblock North-Holland, Amsterdam, 1962.

\bibitem{David2007}
David~D. Hackney.
\newblock Processive motor movement.
\newblock {\em Science}, 316:58--59, 2007.

\bibitem{Anneka2007}
Anneka~M. Hooft, Erik~J. Maki, Kristine~K. Cox, and Josh~E. Baker.
\newblock An accelerated state of myosin-based actin motility.
\newblock {\em Biochemistry}, 46:3513--3520, 2007.

\bibitem{Howard2001}
J.~Howard.
\newblock {\em Mechanics of Motor Proteins and the Cytoskeleton}.
\newblock Sinauer Associates and Sunderland, MA, 2001.

\bibitem{Keller2000}
David Keller and Carlos Bustamante.
\newblock The mechanochemistry of molecular motors.
\newblock {\em Biophysical Journal}, 78:541--556, 2000.

\bibitem{Kudo1990}
S.~Kudo, Y.~Magariyama, and S.~Aizawa.
\newblock Abrupt changes in flagella rotation observed by laser dark-filed
  microscopy.
\newblock {\em Nature}, 346:677--680, 1990.

\bibitem{Kull1996}
F.J. Kull, E.P. Sablin, R.J. Fletterick, and R.D. Vale.
\newblock Crystal structure of the kinesin motor domain reveals a structural
  similarity to myosin.
\newblock {\em Nature}, 380:550--555, 1996.

\bibitem{Liepelt2007}
Steffen Liepelt and Reinhard Lipowsky.
\newblock Kinesin¡¯s network of chemomechanical motor cycles.
\newblock {\em Physical Review Letters}, 98:258102, 2007.

\bibitem{Linden2007}
Martin Lind\'{e}n and Mats Wallin.
\newblock Dwell time symmetry in random walks and molecular motors.
\newblock {\em Biophysical Journal}, 92:3804--3816, 2007.

\bibitem{Masayoshi2002}
Masayoshi Nishiyama, Hideo Higuchi, and Toshio Yanagida.
\newblock Chemomechanical coupling of the forward and backward steps of single
  kinesin molecules.
\newblock {\em Nature Cell Biology}, 4:790--797, 2002.

\bibitem{Noji1997}
H.~Noji, R.~Yasuda, M.~Yoshida, and K.~Kinosita~Jr.
\newblock Direct observation of the rotation of F1-ATPase.
\newblock {\em Nature}, 386:299--302, 1997.

\bibitem{Parmeggiani}
Andrea Parmeggiani, Frank Julicher, Armand Ajdari, and Jacques
Prost.
\newblock Energy transduction of isothermal ratchets: Generic aspects and
  specific examples close to and far from equilibrium.
\newblock {\em Physical Review E}, 60:2127, 1999.

\bibitem{parrondo}
J.M.R. Parrondo and B.J.~De Cisneros.
\newblock Energetics of brownian motors: a review.
\newblock {\em Appl. Phys. A}, 75:179--191, 2002.


\bibitem{Qian2000}
Hong Qian.
\newblock The mathematical theory of molecular motor movement and
  chemomechanical energy transduction.
\newblock {\em Journal of Mathematical Chemistry}, 27:219--234, 2000.

\bibitem{Qian2004}
Hong Qian.
\newblock Motor protein with nonequilibrium potential: Its thermodynamics and
  efficiency.
\newblock {\em Physical Review E}, 69:012901, 2004.

\bibitem{Reimann2002}
Peter Reimann.
\newblock Brownian motors: noisy transport far from equilibrium.
\newblock {\em Physics Reports}, 361:57, 2002.

\bibitem{Feynman1963}
M.~Sands R.P.~Feynman, R.B.~Leighton.
\newblock {\em The Feynman Lectures on Physics}.
\newblock Addison Wesley, Reading, MA, 1963.

\bibitem{Sakakibara1999}
H.~Sakakibara, H.~Kojima, Y.~Sakai, E.~Katayama, and K.~Oiwa.
\newblock Inner-arm dynein c of chlamydomonas flagella is a single-headed
  processive motor.
\newblock {\em Nature}, 400:596--589, 1999.

\bibitem{Katsuyuki2007}
Katsuyuki Shiroguchi and Kazuhiko~Kinosita Jr.
\newblock Myosin v walks by lever brownian motion.
\newblock {\em Science}, 316:1208--1212, 2007.

\bibitem{voboda1994}
K.~Svoboda and S.M. Block.
\newblock Force and velocity measured for single kinesin molecules.
\newblock {\em Cell}, 77:773--784, 1994.

\bibitem{Yuichi2005}
Yuichi Taniguchi, Masayoshi Nishiyama, Yoshiharu Ishhi, and Toshio
Yanagida.
\newblock Entropy rectifies the brownian step of kinesin.
\newblock {\em Nature Chemical Biology}, 1:342--347, 2005.

\bibitem{Vale1996}
R.D. Vale, T.~Funatsu, D.W. Pierce, L.~Romberg, Y.~Harada, and
T.~Yanagida.
\newblock Direct observation of single kinesin molecules moving along
  microtubules.
\newblock {\em Nature}, 380:451--453, 1996.

\bibitem{Vale2000}
Ronald~D. Vale.
\newblock The way things move: Looking under the hood of molecular motor
  proteins.
\newblock {\em Science}, 288:88, 2000.

\bibitem{Vale2003}
Ronald~D. Vale.
\newblock The molecular motor toolbox for intracellular transport.
\newblock {\em Cell}, 112:467--480, 2003.

\bibitem{Wang2005}
Hongyun Wang.
\newblock Chemical and mechanical efficiencies of molecular motors and
  implications for motor mechanisms.
\newblock {\em J. Phys.: Condens. Matter}, 17:S3997--S4014, 2005.

\bibitem{Wang2002}
Hongyun Wang and G.~Oster.
\newblock The stokes efficiency for molecular motors and its applications.
\newblock {\em Europhysics Letters}, 57:134--140, 2002.


\bibitem{Zhang2008}
Yunxin Zhang.
\newblock Three phase model of the processive motor protein kinesin.
\newblock {\em Biophysical Chemistry}, 136:19--22, 2008.

\bibitem{Wang2007}
Wenwei Zheng, Dagong~Fan, Zhisong~Wang, Min~Feng.
\newblock Kinesin is an evolutionarily fine-tuned molecular ratchet-and-pawl
  device of decisively locked direction.
\newblock {\em Biophys. J.}, 93:3363--3372, 2007.

\end{thebibliography}
\end{document}